\documentclass[aps,prb,twocolumn,superscriptaddress,showpacs]{revtex4} 

\usepackage[dvips]{graphicx}
\usepackage{bm}
\usepackage{latexsym} 
\usepackage{amsmath}
\usepackage{amssymb}

\newcommand{\ui}{{\rm i}} 

\newcommand{\bmr}{{\bm r}}

\newcommand{\bmS}{{\bm S}}

\newcommand{\bmq}{{\bm q}}

\newcommand{\bmk}{{\bm k}}

\newcommand{\kB}{k_{\rm B}} 

\newcommand{\bmM}{{\bm M}}
\newcommand{\bmm}{{\bm m}}
\newcommand{\bmz}{{\bm z}}
\newcommand{\bmH}{{\bm H}}
\newcommand{\bmE}{{\bm E}}
\newcommand{\bmJ}{{\bm J}}

\newcommand{\jump}[1]{\ensuremath{[\![#1]\!]} }

\begin{document}

\title{
Spin Seebeck effect in antiferromagnets and compensated ferrimagnets
}

\author{Yuichi Ohnuma}
\email{y-ohnuma@imr.tohoku.ac.jp}
\affiliation{Institute for Materials Research, Tohoku University, Sendai 980-8577, Japan}

\author{Hiroto Adachi}
\affiliation{Advanced Science Research Center, Japan Atomic Energy Agency, Tokai 319-1195, Japan}
\affiliation{CREST, Japan Science and Technology Agency, Sanbancho, Tokyo 102-0075, Japan}

\author{Eiji Saitoh}
\affiliation{Institute for Materials Research, Tohoku University, Sendai 980-8577, Japan}
\affiliation{Advanced Science Research Center, Japan Atomic Energy Agency, Tokai 319-1195, Japan}
\affiliation{CREST, Japan Science and Technology Agency, Sanbancho, Tokyo 102-0075, Japan}
\affiliation{WPI Research Center, Advanced Institute for Material Research, Tohoku University, Sendai 980-8577, Japan}

\author{Sadamichi Maekawa}
\affiliation{Advanced Science Research Center, Japan Atomic Energy Agency, Tokai 319-1195, Japan}
\affiliation{CREST, Japan Science and Technology Agency, Sanbancho, Tokyo 102-0075, Japan}

\date{\today}

\begin{abstract} 
We theoretically investigate the spin Seebeck effect (SSE) in antiferromagnets and ferrimagnets, and show that the SSE vanishes in antiferromagnets but survives in ferrimagnets even at the magnetization compensation point despite the absence of its saturation magnetization. The non-vanishing SSE in ferrimagnets stems from two non-degenerate magnons. We demonstrate that the magnitude of the SSE in ferrimagnets is unchanged across the magnetization compensation point. 
\end{abstract}



\maketitle 

\section{Introduction~\label{Sec:Intro}} 

Much attention is now focused on the thermal effects in spintronics, and the emergent research field of spin caloritronics is rapidly developing.~\cite{SpinCalo,Bauer12} One of the most important issues in spin caloritronics is the spin Seebeck effect (SSE).~\cite{Uchida08} The SSE is the mechanism by which a spin voltage is generated from a temperature gradient in a magnetic material over a macroscopic scale of several millimeters.~\cite{Uchida12} Because the spin voltage is a potential for electron spins to drive spin currents, this spin voltage injects a pure spin current, i.e., a spin polarized current which is unaccompanied by a charge current, from the ferromagnet into an attached nonmagnetic metal. The inverse spin Hall effect (ISHE)~\cite{Saitoh06,Valenzuela06} converts the injected spin current into a transverse electric voltage and hence the SSE is electrically detectable. Since its discovery in 2008, this phenomenon has drawn much interest as a simple way of generating pure spin currents that are needed for future spin-based technology,~\cite{Zutic04,Kirihara12} and the recent observation of the giant SSE in InSb~\cite{Jaworski12} has attracted a considerable attention. 
 
The SSE has been observed in various ferromagnetic materials ranging from metallic ferromagnets, Ni$_{81}$Fe$_{19}$~\cite{Uchida08} and Co$_2$MnSi,~\cite{Bosu11} to semiconducting ferromagnet (Ga,Mn)As,~\cite{Jaworski10,Jaworski11} to insulating magnets LaY$_2$Fe$_5$O$_{12}$~\cite{Uchida10a} and (Mn,Zn)Fe$_2$O$_4$.~\cite{Uchida10c} Although LaY$_2$Fe$_5$O$_{12}$ and (Mn,Zn)Fe$_2$O$_4$ are classified into ferrimagnets in a rigorous terminology, the current understanding of the SSE in these systems relies on a modeling as ferromagnets~\cite{Xiao10,Adachi11} because the low-energy magnetic properties relevant to the SSE are well described by a ferromagnet modeling owing to the large gap between the acoustic and optical magnons. These observations have established the SSE as a universal aspect of ferromagnets.

Besides ferromagnets, ferrimagnets and antiferromagnets are known as prototypes of magnetic materials.~\cite{Kittel-intro} A ferrimagnet is an ordered spin system in which two sublattice magnetizations point in the opposite directions, and an antiferromagnet is classified as a special case of a ferrimagnet for which both sublattices have equal saturation magnetizations. Recently, there has been an on-going attempt to develop antiferromagnetic metal spintronics, and several experimental~\cite{Park11} and theoretical~\cite{Shick10,Swaving11,Hals11} work are already in progress. Regarding ferrimagnets, the intriguing characteristics of ferrimagnetic ordering are now drawing considerable attention~\cite{Stanciu07,Radu11} in developing a ultrafast magnetization manipulation technique. Therefore, it is quite natural to ask whether the SSE can be observed in antiferromagnets and ferrimagnets. 

In this paper, we address the issue of observing the SSE in antiferromagnets and ferrimagnets. Especially, we focus on the SSE in ferrimagnets with magnetization compensation. A certain class of ferrimagnets are known to possess a magnetization compensation temperature $T_{\rm M}$ (angular-momentum compensation temperature $T_{\rm A}$), at which the two sublattice magnetizations (spins) have the same magnitudes but opposite directions, leading to net-zero saturation magnetization (spin angular momentum).~\cite{Neel64,Pauthenet58,Geller65,LeCraw65,Chikazumi-text} We show that two non-degenerate magnons give rise to the non-vanishing SSE at $T_{\rm M}$ or $T_{\rm A}$ despite the absence of net saturation magnetization or total spin. Also, we show that for a uniaxial antiferromagnet the SSE vanishes because the thermal spin injection by the two degenerate magnons is perfectly compensated. Moreover, the SSE in an easy-plane antiferromagnet is shown to disappear because in this instance neither magnon carries spins. 

This paper is organized as follows. In Sec.~II, we investigate the SSE in uniaxial antiferromagnets as well as ferrimagnets with magnetization compensation. Next, in Sec.~III we discuss the SSE in easy-plane antiferromagnets. Finally, in Sec.~IV we summarize and discuss our results. 

\section{Spin Seebeck effect in uniaxial antiferromagnets and ferrimagnets \label{Sec:Ferri}} 

\begin{figure}[t] 
  \begin{center}
    \scalebox{0.13}[0.13]{\includegraphics{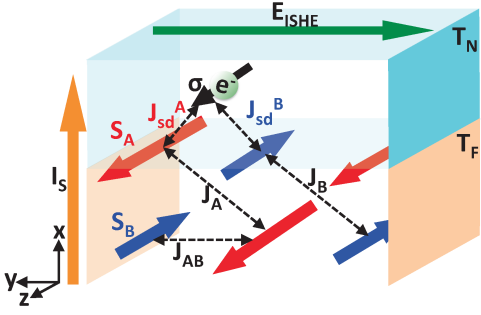}}
  \end{center}
\caption{ 
(Color online) Schematic view of a hybrid structure composed of a nonmagnetic metal ($N$) and a ferrimagnet ($F$) with two sublattices $A$ and $B$. 
}
\label{fig1_Ohnuma}
\end{figure}

As a general model of ferrimagnets and antiferromagnets, we consider the following Hamiltonian defined on a lattice composed of two sublattices $A$ and $B$,~\cite{Wolf61} 
\begin{eqnarray}
  {\cal H}_{F} &=& 
  -J_{A} \sum_{\langle i,i' \rangle \in A} \bmS_{A,i} \cdot \bmS_{A,i'} 
  -J_{B} \sum_{\langle j,j' \rangle \in B} \bmS_{B,j} \cdot \bmS_{B,j'} \nonumber \\
  &+& J_{AB} \sum_{\langle i \in A, j \in B \rangle } \bmS_{A,i} \cdot \bmS_{B,j} 
  + \delta {\cal H}_A + \delta {\cal H}_B, 
\label{Eq:H-ferri01}
\end{eqnarray}
where $J_{A}$ and $J_B$ ($J_{AB}$) are the nearest-neighbor intra-sublattice (inter-sublattice) exchange integrals, and $\langle , \rangle$ specifies nearest neighbor bonding (see Fig.~\ref{fig1_Ohnuma}). The last two terms in Eq.~(\ref{Eq:H-ferri01}) for sublattice $L=A,B$ are given by $\delta {\cal H}_L = \sum_{ i \in L} [g_{_L} \mu_0 {\bm H}_0 \cdot \bmS_{L,i} -\frac{D_L}{2} (\widehat{\bm z} \cdot \bmS_{L,i})^2]$, where $\mu_0$ is the Bohr magneton, ${\bm H}_0 = -H_0 \widehat{\bm z}$ is the external magnetic field, $g_{_L}$ and $D_L$ are the effective g-factor and the anisotropy constant for sublattice $L$. 

First, we use the spin-wave approximation to diagonalize Eq.~(\ref{Eq:H-ferri01}). Following standard procedures~\cite{Kittel-text} using the linear Holstein-Primakoff transformation for spin operators $S_{L,i}^\pm = S_{L,i}^{x} \pm \ui S_{L,i}^{y} $ ($L=A,B$), the Hamiltonian (\ref{Eq:H-ferri01}) is diagonalized to be 
\begin{equation}
  {\cal H}_{F} 
  = 
  \hbar \sum_\bmq \Big( \omega^{+}_\bmq \alpha^\dag_\bmq \alpha_\bmq + 
  \omega^{-}_\bmq \beta^\dag_\bmq \beta_\bmq \Big), 
\label{Eq:H-ferri02}
\end{equation}
where $\omega^{\pm}_\bmq= \frac{1}{2}
\sqrt{(\epsilon^{A}_\bmq+\epsilon^{B}_\bmq)^2 - 4 \eta_\bmq^2 
} \pm (\epsilon^{A}_\bmq- \epsilon^{B}_\bmq)$, 
and the precise forms of $\epsilon^{A}_\bmq$, $\epsilon^{B}_\bmq$, and $\eta_\bmq$ 
are given by 
$\epsilon^A_\bmq= 2 z_{0} J_{A} S_A [1-\gamma_\bmq] + z_0 J_{AB} S_B + (g^{_A} \mu_0 H_0 + D_A S_A)$ 
and $\epsilon^B_\bmq= 2 z_{0} J_{B} S_B[1-\gamma_\bmq] + z_0 J_{AB} S_A + (-g_{_B} \mu_0 H_0 + D_B S_B)$. 
Here, $\gamma_\bmq= z_0^{-1} \sum_{\bm \delta} e^{\ui \bmq \cdot {\bm \delta}}$ is defined by the sum over $z_0$ nearest neighbors of the original lattice, and $\eta_\bmq= J_{AB} \sqrt{S_A S_B} \sum_{\bm \delta'}e^{\ui \bmq \cdot {\bm \delta'}}$ is defined by the sum over $z_0$ nearest neighbors of the sublattice $A$ or $B$. In this paper, we assume a cubic lattice for simplicity. In Eq.~(\ref{Eq:H-ferri02}), the magnon operators $\alpha_\bmq$ and $\beta_\bmq$ are defined by the Bogoliubov transformation~\cite{Sparks-text} 
$a_\bmq = u^+_\bmq \alpha_\bmq + u^-_\bmq \beta^\dag_\bmq$ 
and 
$b_\bmq = u^-_\bmq \alpha^\dag_\bmq + u^+_\bmq \beta_\bmq$, 
where 
and $a_\bmq$ and $b_\bmq$ are the Fourier transforms of operators 
$a_i= (2 S_A)^{-\frac{1}{2}}S^+_{A,i}$ and $b_i= (2 S_B)^{-\frac{1}{2}}S^-_{B,i}$ 
with $S_A=|\bmS_A|$ and $S_B = |\bmS_B|$, 
and ${u^+_\bmq}^2- {u^-_\bmq}^2=1$. 

\begin{figure}[t] 
  \begin{center}
    \scalebox{0.22}[0.22]{\includegraphics{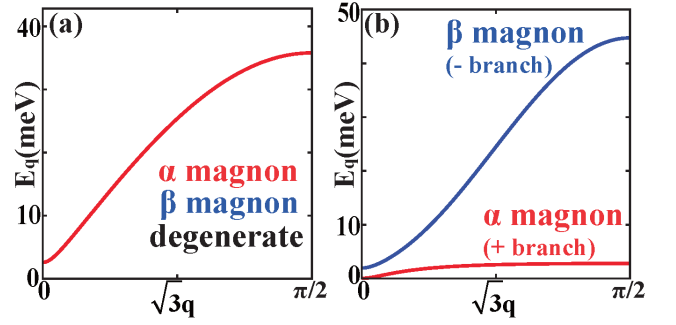}}
  \end{center}
\caption{ 
(Color online) Spin-wave spectra ($H_0=0$) with $\bmq$ along the [111] direction calculated from Eq.~(\ref{Eq:H-ferri02}) using parameters for (a) a uniaxial antiferromagnet NiO, and (b) a compensated ferrimagnet Er$_3$Fe$_5$O$_{12}$. The wavevector $q$ is measured in units of the inverse of the nearest-neighbor distance. 
} 
\label{fig2_ohnuma}
\end{figure}

In Fig.~\ref{fig2_ohnuma}, the spin-wave spectra ($H_0=0$) calculated from Eq.~(\ref{Eq:H-ferri02}) for a uniaxial antiferromagnet NiO and a compensated ferrimagnet Er$_3$Fe$_5$O$_{12}$ are plotted. For NiO, we use $J_{AB}=6.3$~meV ($J_A=J_B=0$), $D=0.1$~meV, $S_A=S_B=0.92$,~\cite{Chikazumi-text,Hutchings72} whereas for Er$_3$Fe$_5$O$_{12}$, we assign the net spin of the rare-earth ions (the ferric ions) to $S_A$ ($S_B$) on a model cubic lattice, and we set $J_A=0$~meV, $J_{B}=0.68$~meV, $J_{AB}=0.19$~meV, $S_{A}=4.2$, $S_{B}=2.5$, $g_A=1.4$, $g_B=2.0$ $D_{A}=3.5 \times 10^{-3}$~meV, and $D_{B}=3.0\times10^{-4}$~meV to reproduce the N\'{e}el temperature $T_{\rm N \text{\'{e}} el } =556$~K and the magnetization-compensation temperature $T_{\rm M}=83$~K.~\cite{Chikazumi-text,Pearson62} As is well known, the two antiferromagnetic magnons are degenerate if $H_0=0$, whereas the two ferrimagnetic magnons are non-degenerate because of the inequivalence of the two sublattices.

We discuss now the SSE in uniaxial antiferromagnets and ferrimagnets modeled by Eq.~(\ref{Eq:H-ferri01}). Note that a uniaxial antiferromagnet can be modeled as a special case of a ferrimagnet. We consider a model shown in Fig.~\ref{fig1_Ohnuma}, in which a ferrimagnet ($F$) and a nonmagnetic metal ($N$) are interacting weakly through the $s$-$d$ exchange interaction at the interface. We assume that the ferrimagnet $F$ has a local temperature $T_F$, and the nonmagnetic metal $N$ has a local temperature $T_N$. We analyze the SSE in the longitudinal configuration~\cite{Uchida10b} by employing the linear-response formulation of the SSE in a ferromagnet developed in Refs.~\onlinecite{Adachi11} and~\onlinecite{Adachi12c}. The $s$-$d$ interaction at the interface is modeled by 
\begin{eqnarray}
  {\cal H}_{\rm sd} &=& 
 \sum_{i,j \in F/N\mathchar`-\text{interface}} 
  \Big( 
  J^A_{\rm sd} {\bm \sigma}_i \cdot \bmS_{A,i} 
  + J^B_{\rm sd} {\bm \sigma}_j \cdot \bmS_{B,j}
  \Big), 
  \label{Eq:Hsd01} 
\end{eqnarray} 
where, for sublattice $L=A,B$, $J^{L}_{\rm sd}$ is the $s$-$d$ exchange interaction at the $F/N$ interface, ${\bm \sigma}_i$ is the itinerant spin density operator in $N$. The total Hamiltonian of the system, ${\cal H}$, is then given by 
\begin{equation}
  {\cal H} = {\cal H}_F+ {\cal H}_N+ {\cal H}_{\rm sd}, 
  \label{H-total01}
\end{equation}
where ${\cal H}_{N}$ is the single-particle Hamiltonian of the conduction electrons in $N$ (see, e.g., Eq.~(31) in Ref.~\onlinecite{Takahashi08}). 

\begin{figure}[t] 
  \begin{center}
    \scalebox{0.4}[0.4]{\includegraphics{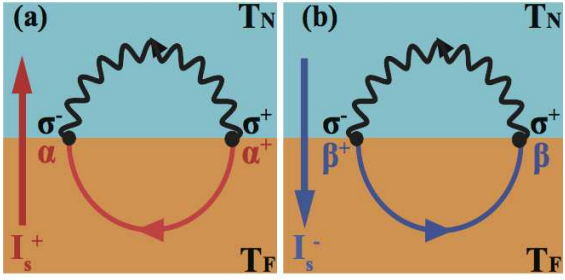}}
  \end{center}
\caption{ 
(Color online) 
Feynman diagram representing two processes relevant to the SSE in uniaxial antiferromagnets and ferrimagnets. (a) Spin current injected by $\alpha_\bmq$ magnons ($I_s^+$). (b) Spin current injected by $\beta_\bmq$ magnons ($I_s^-$). The signs of $I_s^+$ and $I_s^-$ are opposite. The solid and wavy lines represent magnon and itinerant spin-density propagators, respectively. 
}
\label{fig3_Ohnuma}
\end{figure}

The central quantity that characterizes the SSE is the spin current $I_s$ injected into $N$, because it is proportional to the experimentally detectable electric field $\bmE_{\rm ISHE}$ via ISHE:~\cite{Saitoh06,Valenzuela06} 
\begin{equation}
  \bmE_{\rm ISHE} = \theta_{\rm SH} \rho 
  \bmJ_s 
  \times {\bm \sigma}, 
\end{equation}
where $\theta_{\rm SH}$ and $\rho$ are respectively the spin-Hall angle and the resistivity of $N$, and $\bmJ_s= (I_s/A_{\rm int})\widehat{\bm x}$ is the spin-current density across the $F/N$ interface having a contact area $A_{\rm int}$. Following Refs.~\onlinecite{Adachi11} and~\onlinecite{Adachi12c}, we calculate $I_s$ as the rate of change of the spin accumulation in $N$, i.e., $I_s= \sum_{i \in N} \langle \partial_t \sigma_i^z \rangle$ where $\langle \cdots \rangle$ denotes the statistical average. 
What is special in the present calculation is that we need to express the $s$-$d$ interaction [Eq.~(\ref{Eq:Hsd01})] in terms of the $\alpha_\bmq$ and $\beta_\bmq$ operators [``$\pm$'' branches in Eq.~(\ref{Eq:H-ferri02})], because these are the magnon operators in $F$. Following procedures presented in Appendix~\ref{Sec:Append01}, the spin current injected in $N$ is expressed as 
\begin{eqnarray}
  I_s &=& -\frac{2\sqrt{2}}{\sqrt{N_{N}N_{F}}\hbar} 
  {\rm Re} \sum_{\bmk, \bmq}\int_{\omega} 
  \Big[{\cal J}^+_{\rm sd}(\bmk,\bmq) {\cal A}^{K}_{\bmk,\bmq}(\omega) \nonumber \\
    && \qquad + 
    {\cal J}^-_{\rm sd}(\bmk,\bmq) {\cal B}^{K}_{\bmk,\bmq}(\omega) 
    \Big], 
\label{Eq:Is-total02}
\end{eqnarray}
where ${\cal A}^{K}$ (${\cal B}^{K}$) is the Keldysh component of the interface correlation function between magnon operator $\alpha_\bmq$ ($\beta^\dag_\bmq$) and the itinerant spin-density operator $\sigma^-_\bmk$ (see Appendix~\ref{Sec:Append01}), and we have introduced the shorthand notation $\int_\omega=\int_{-\infty}^\infty \frac{d \omega}{2 \pi}$. Here ${\cal J}^\pm_{\rm sd}(\bmk,\bmq)$ is the effective $s$-$d$ interaction written in terms of magnon operators, and the precise definition is given in the Appendix~\ref{Sec:Append01}. 

We perform the perturbative approach in term of the $s$-$d$ interaction at the interface to evaluate Eq.~(\ref{Eq:Is-total02}). Then, the spin current $I_s$ is given by the two diagrams shown in Fig.~\ref{fig3_Ohnuma}, and accordingly, $I_s$ has two terms: 
\begin{equation}
  I_s = I^+_s + I_s^-, 
\label{Eq:Is-total01}
\end{equation}
where $I_s^\pm$, representing the contribution from the $\pm$ branch, is expressed by 
\begin{eqnarray}
  I_s^\pm &=& 
  \pm \sum_{\bmk,\bmq} 
  \frac{ 8 N_{\rm int} \jump{|{\cal J}_{\rm sd}^\pm(\bmk,\bmq)|^2} }{N_N N_F\hbar^2} 
    \int_\omega 
   {\rm Im} \chi^R(\bmk,\omega) \nonumber \\
  && \times    {\rm Im} {\cal G}^R_\pm (\bmq,\omega) 
  \Big[ \coth(\tfrac{\hbar \omega}{2 \kB T_{N}})  
    - \coth(\tfrac{\hbar \omega}{2 \kB T_{F}})   \Big]. 
\label{Eq:Is-total03}
\end{eqnarray}  
Here $N_{\rm int}$ is the number of localized spins at the $F$/$N$ interface, $N_N$ ($N_{F}$) is the number of lattice sites in $N$ (sublattice sites in $F$), 
$\jump{|{\cal J}_{\rm sd}^\pm(\bmk,\bmq)|^2} = S_A (J^{A}_{\rm sd} u^\pm_\bmq)^2+ S_B (J^{B}_{\rm sd} u^\mp_\bmq)^2 $. In Eq.~(\ref{Eq:Is-total03}), 
$\chi^R(\bmk,\omega)=\chi_N/(1+\lambda_N^2 k^2 - \ui \omega \tau_{\rm sf})$ 
where $\chi_N$, $\lambda_N$, $\tau_{\rm sf}$ are respectively the paramagnetic susceptibility, the spin-diffusion length, and the spin-flip relaxation time in $N$, 
and 
${\cal G}_\pm^R(\bmq,\omega)= 
1/(\omega - \omega^\pm_\bmq+ \ui \alpha_\pm \omega)$ 
where 
$\alpha_\pm$ is the damping parameter in $F$. Note that the signs of the spin current injected by the $\alpha_\bmq$ magnons ($I_s^+$) and that by the $\beta_\bmq$ magnons ($I_s^-$) are opposite. 
 
We first consider the SSE in a uniaxial antiferromagnet. As is depicted in Fig.~\ref{fig2_ohnuma} (a), the two magnons in a uniaxial antiferromagnet are degenerate if $H_0=0$. Moreover, owing to the equivalence of sublattices $A$ and $B$, the $s$-$d$ exchange interactions at the interface for these two sublattices are the same ($J_{\rm sd}^{A} = J_{\rm sd}^{B}$). From these conditions we obtain $|I_s^+| = |I_s^-|$ resulting in a null SSE due to Eq.~(\ref{Eq:Is-total01}), i.e., $I_s=0$. Thus, the SSE vanishes in a uniaxial antiferromagnet under a negligibly small external magnetic field because of the perfect compensation of the spin injection by the two degenerate magnons.

\begin{figure}[t] 
  \begin{center} 
    \scalebox{0.33}[0.33]{\includegraphics{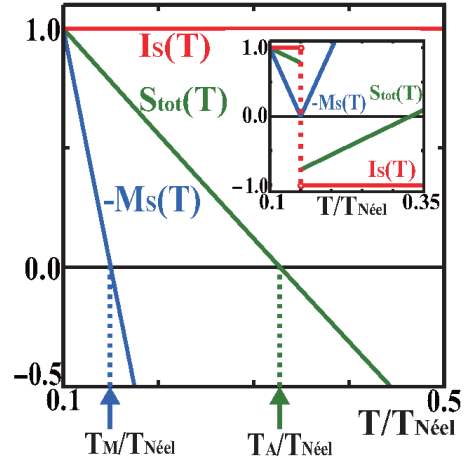}}
  \end{center}
\caption{ (Color online) 
Temperature dependence of the SSE signal $I_s$ [red, Eq.~(\ref{Eq:Is-total01})], saturation magnetization $M_s$ [blue, Eq.~(\ref{Eq:Ms01})], and total angular momentum $S_{\rm tot}$ [green, Eq.~(\ref{Eq:Stot01})], calculated for a compensated ferrimagnet Er$_3$Fe$_5$O$_{12}$ using the same parameters as in Fig.~\ref{fig2_ohnuma} (b). The case for a $M_s$ pinned by the anisotropy field is shown; the data is normalized by its value at $T/T_{\rm N \text{\'{e}}el}=0.1$. Inset: The case for a $M_s$ pinned by the external magnetic field is shown. 
}
\label{fig4_Ohnuma}
\end{figure}

We next consider the SSE in a ferrimagnet close to the magnetization compensation point, in which the two magnons are no longer degenerate. 
Figure~\ref{fig4_Ohnuma} shows the temperature dependence of the SSE signal $I_s(T)$ calculated from Eqs.~(\ref{Eq:Is-total01}) and
~(\ref{Eq:Is-total03}) 
for a compensated ferrimagnet Er$_3$Fe$_5$O$_{12}$ by using the same parameters as in Fig.~\ref{fig2_ohnuma} (b). In Fig.~\ref{fig4_Ohnuma} we also plot the saturation magnetization 
\begin{equation}
  {M_{\rm s}} = {\mu_{0}} \Big( \frac{g_{A}}{N_{F}}\sum_{i \in A} 
  \langle S^z_{A,i} \rangle + \frac{g_{B}}{N_{F}}\sum_{j \in B} 
  \langle S^z_{B,j} \rangle \Big) 
\label{Eq:Ms01}
\end{equation}
to determine the magnetization compensation point defined by $M_{\rm s}(T_{\rm M})=0$. 
In addition, we plot the total angular momentum 
\begin{equation}
  S_{\rm tot}= \langle S^z \rangle 
  \label{Eq:Stot01} 
\end{equation}
to determine the angular-momentum compensation point defined by $S_{\rm tot}(T_{\rm A})=0$. 
Here, $S^z$ is the $z$-component of the total spin $\bmS= \bmS_A + \bmS_B$, i.e., 
\begin{equation}
  S^z = \frac{1}{N_{F}}\sum_{i \in A} S^z_{A,i} 
  + \frac{1}{N_{F}}\sum_{j \in B} S^z_{B,j}. 
  \label{Eq:Sztot00}
\end{equation}
Clearly we see that the SSE signal is unchanged across both compensation points, either $T_{\rm M} \approx 0.15 T_{\rm N\text{\'{e}}el}$ or $T_{\rm A} \approx 0.32 T_{\rm N\text{\'{e}}el}$. We performed the same calculation for several different choices of parameters, and confirmed that the SSE is unchanged across $T_{\rm M}$ and $T_{\rm A}$. 

\section{Spin Seebeck effect in easy-plane antiferromagnets \label{Sec:Easy}}  

In this section, we show that the SSE in easy-plane antiferromagnets vanishes under a zero magnetic field because neither of magnons carries spins in easy-plane antiferromagnets. We consider the following Hamiltonian for easy-plane antiferromagnets:~\cite{Maekawa73} 
\begin{eqnarray}
  {\cal H}_{eAF} &=& 
  J \sum_{\langle i \in A, j \in B \rangle } \bmS_i^{_A} \cdot \bmS_{j}^{_B} \nonumber \\
  &+& \sum_{L=A, B} \sum_{ i \in L} [g \mu_0 \bmH_0 \cdot \bmS^{_L}_i  -\frac{D}{2} 
(\widehat{\bm z} \cdot \bmS_i^{_L})^2]
\label{Eq:H-eAF01}
\end{eqnarray}
where $J$  is the nearest-neighbor exchange integrals, ${\bm H}_0 = H_0 \widehat{\bm x}$ is the external magnetic field, and $g$ is the g-factor and $D<0$ is the anisotropy constant which selects the $x$-$y$ plane as an easy plane. Note that the external magnetic field is applied along the $x$ axis, and we assume $\bmS_A \parallel \widehat{\bm z}$ and $\bmS_B \parallel -\widehat{\bm z}$ when $H_0=0$. 
Following Ref.~\onlinecite{Maekawa73}, we introduce the linear Holstein-Primakoff transformation by choosing the direction of each canted sublattice spin in the ground state as a spin quantizing axis. Performing a $\pi/4$ rotation to the operators to separate the mixing of the two spin operators and using the Bogoliubov transformation, Eq.~(\ref{Eq:H-eAF01}) is diagonalized to be  
\begin{equation}
  {\cal H}_{eAF} 
  = 
  \hbar \sum_\bmq \Big( \varepsilon^{+}_\bmq \xi^\dag_\bmq \xi_\bmq + 
  \varepsilon^{-}_\bmq \zeta^\dag_\bmq \zeta_\bmq \Big), 
\label{Eq:H-eAF02}
\end{equation}
where $ \varepsilon^{\pm}_{\bmq} = \sqrt{(A^{\pm}_{\bmq} + 2 B^{\pm}_{\bmq})(A^{\pm}_{\bmq} - 2 B^{\pm}_{\bmq})} $, $ A^{\pm}_{\bmq} = 2 z_0 JS \cos 2\theta + g\mu_{0} H_{0} \sin\theta + |D|S \mp z_0JS (\cos 2\theta - 1)\gamma_{\bmq} $, $ B^{\pm}_{\bmq} = \mp z_0 JS (\cos 2\theta - 1)\gamma_{\bmq} - |D|S/2 $,
and $\gamma_\bmq= z_0^{-1} \sum_{\bm \delta} e^{\ui \bmq \cdot {\bm \delta}}$ 
is defined by the sum over $z_0$ nearest neighbors. 
In the above equation, $\theta$ is the canted angle of the sublattice magnetization, and the magnon operators $\xi_\bmq$ and $\zeta_\bmq$ are defined by the Bogoliubov transformation $\frac{1}{\sqrt{2}} \big( a_\bmq + b_{-\bmq} \big) = 
  u_\bmq \xi_\bmq + v_\bmq \xi^\dag_{-\bmq}$ and $\frac{1}{\sqrt{2}} \big( a_\bmq - b_{-\bmq} \big)  =    x_\bmq \zeta_\bmq + y_\bmq \zeta^\dag_{-\bmq}$, 
where $u_\bmq^2- v_\bmq^2=1$ and $x_\bmq^2- y_\bmq^2=1$ are real coefficients, and 
$a_\bmq$ and $b_\bmq$ are the Fourier transforms of Holstein-Primakoff operators $a_i$ and $b_i$. As is seen in Fig.~3 of Ref.~\onlinecite{Maekawa73}, the two magnons in the easy-plane antiferromagnet are not degenerate even when $H_0=0$. 

Now we discuss the SSE in an easy-plane antiferromagnet modeled by Eq.~(\ref{Eq:H-eAF01}) by using the same procedure as in the previous section. 
As before, we consider a system in which an easy-plane antiferromagnet ($eAF$) having local temperature $T_{eAF}$ and a nonmagnetic metal ($N$) having local temperature $T_N$ are interacting weakly through the $s$-$d$ exchange interaction at the interface. In the absence of an external magnetic field, a direct calculation shows that the spin current $I_s$ injected into $N$ is identically zero, i.e., 
\begin{equation}
  I_s = 0. 
  \label{Eq:eAF-SSE00}
\end{equation}
This is understood by investigating the $z$-component of the total spin $\bmS=\bmS_A+\bmS_B$ [Eq.~(\ref{Eq:Sztot00})]. In the case of a uniaxial antiferromagnet discussed in the previous section, the expectation value of $S^z$ is given by 
\begin{equation}
  \langle S^z \rangle = \Big( S_A - \frac{1}{N_{F}}\sum_\bmq \langle \alpha^\dag_\bmq \alpha_\bmq \rangle \Big)  
  - \Big( S_B - \frac{1}{N_{F}} \sum_\bmq \langle \beta^\dag_\bmq \beta_\bmq \rangle \Big),  
  \label{Eq:Sztot01}
\end{equation}
where $\alpha_\bmq$ and $\beta_\bmq$ are the magnon operators defined in Sec.~\ref{Sec:Ferri}. 
Equation~(\ref{Eq:Sztot01}) means that the $\alpha_\bmq$ magnons carries spin one while $\beta_\bmq$ magnon carries spin minus one in a uniaxial antiferromagnet. On the other hand, the expectation value of $S^z$ in an easy-plane antiferromagnet under discussion is calculated to be identically zero, i.e., 
\begin{equation}
  \langle S^z \rangle= 0. 
  \label{Eq:Sztot02}
\end{equation}
Equation~(\ref{Eq:Sztot02}) means that magnons in an easy-plane antiferromagnet are similar to a linearly-polarized photon and hence neither of magnons carries spins if $H_0=0$. 

In the presence of a finite external magnetic field $\bmH_0=H_0 \widehat{\bm x}$ with a nonzero canted angle $\theta$, however, the $x$-component of $\bmS$ becomes nonzero. In this situation, the spin current $I_s$ injected into $N$ is shown to have the polarization along the $x$ axis, and its magnitude is given by 
\begin{equation}
  I_s = \frac{g \mu_0 H_0 }{J z_0 } \left( I^+_s + I_s^- \right), 
\label{Eq:Is-eAF-SSE01}
\end{equation}
where 
\begin{eqnarray}
  I_s^{\pm} &=& \frac{4 J^2_{\rm sd} N_{\rm int}}{N_{N}N_{F} \hbar^2} 
  \sum_{\bmk,\bmq} \int_{\omega} 
      {\rm Im} \chi^R(\bmk,\omega) 
  {\rm Im} {\cal F}^R_{\pm} (\bmq,\omega) \nonumber \\
  && \times \Big[ \coth(\tfrac{\hbar \omega}{2 \kB T_{N}})  
    - \coth(\tfrac{\hbar \omega}{2 \kB T_{eAF}})   \Big]. 
 \label{Eq:Is-eAF-SSE02}
\end{eqnarray}
Here, ${\cal F}_\pm^R(\bmq,\omega)= 1/(\omega - \varepsilon^\pm_\bmq+ \ui \alpha_\pm \omega)$ is the retarded component of the magnon propagator with $\alpha_\pm$ is the damping parameter. Note that the signs of $I^+_s$ and $I^-_s$ are the same, in contrast to the case of a uniaxial antiferromagnet. 

From Eqs.~(\ref{Eq:eAF-SSE00}) and~(\ref{Eq:Is-eAF-SSE01}), we conclude that the SSE vanishes in an easy-plane antiferromagnet if $H_0=0$.

\section{Discussion and Conclusion \label{Sec:Conclusion}}

The main result of this paper is that the SSE in antiferromagnets vanishes, whereas the SSE in ferrimagnets persists and is insensitive to either magnetization or angular-momentum compensation effects. 
The interpretation is as follows. 
For the SSE to occur, the existence of the transverse fluctuations of the total spin, i.e., $S^x_{\rm tot}$ and $S^y_{\rm tot}$, is needed. 
For a ferrimagnet at $T_{\rm M}$ or $T_{\rm A}$, fluctuations of $S^{x,y}$ do not vanish even when $S_{\rm tot} =0$ or $M_{\rm s}=0$. Therefore, ferrimagnetic magnons can always generate the SSE. Only for a uniaxial antiferromagnet, where the two magnons are degenerate, the SSE from the two degenerate magnons with the opposite sense compensates perfectly. Note that neither magnon in an easy-plane antiferromagnet carries spins. 

Our conclusion is not modified by considering the phonon-drag contribution to the SSE~\cite{Adachi10} because, as discussed in Refs.~\onlinecite{Adachi12a} and~\onlinecite{Adachi12b}, phonon drag can be taken into account 
by replacing $T_{F}$ and $T_N$ in Eq.~(\ref{Eq:Is-total03}) with an effective magnon temperature $T^*_F$ and effective spin accumulation temperature $T_N^*$. We also note that the magnon excitations are well defined even at $T_{\rm A}$. The presumed divergence of the magnon damping parameter at $T_{\rm A}$~\cite{Stanciu06} does not exist when we recall the condition justifying the expansion used in Ref.~\onlinecite{Wangsness53}, where such an effect manifests itself as an enhancement of the damping parameter without any divergence (see Appendix~\ref{Sec:Append02}). Note that the magnitude of magnon damping has less effect on the longitudinal SSE, although it has a large influence on the transverse SSE.

To conclude, we have theoretically investigated the SSE in antiferromagnets and ferrimagnets, and shown that the SSE vanishes in antiferromagnets whereas it persists at either the magnetization or the angular-momentum compensation points of ferrimagnets, despite the absence of its saturation magnetization or total spin. 
Because a fringing field by saturation magnetization is suppressed at the magnetization compensation point, this phenomenon can be useful for constructing a pure spin current device which is free from crosstalk of the fringing field. 

\acknowledgments 
We are grateful to K. Uchida. This work is was financially supported by a Grant-in-Aid from MEXT, Japan, and a Fundamental Research Grants from CREST-JST, PRESTO-JST, Japan.

\appendix

\section{Linear-response expression of magnon-driven spin injection in ferrimagnets~\label{Sec:Append01}} 

In this Appendix, we derive Eq.~(\ref{Eq:Is-total02}) in the main text. 
We consider a system described by the Hamiltonian (\ref{H-total01}), and calculate the 
spin current $I_s= \sum_{i \in N} \langle \partial_t \sigma_i^z \rangle$. 
We use the momentum representation of ${\sigma}^z_i$ and calculate the quantity 
$I_s= \sqrt{N_{N}} \langle \partial_{t} \sigma^{z}_{\bmk_{0}} \rangle_{\bmk_{0} \to 0}$. 
The Heisenberg equation of motion for $\sigma^z_{\bmk_0}$ gives 
\begin{eqnarray} 
  \partial_{t} \sigma^{z}_{\bmk_{0}} 
  &=&   \frac{\ui}{\hbar} \sum_{\bmk, \bmq} \Big[ 
    \frac{\sqrt{8S_A}J^A_{\rm sd}(\bmk,\bmq)}{\sqrt{N_{F}}N_N } 
  (u^{+}_{\bmq}\alpha^{-}_{\bmq}+u^{-}_{\bmq}\beta^{+}_{\bmq}) 
  \sigma_{\bmk}^{-} \nonumber \\
  &+& 
  \frac{\sqrt{8S_B}J^B_{\rm sd}(\bmk,\bmq)}{\sqrt{N_{F}}N_N } 
  (u^{+}_{\bmq}\beta^{+}_{\bmq}+u^{-}_{\bmq}\alpha^{-}_{\bmq})
  \sigma_{\bmk}^{-} \Big] 
  +{\rm h.c.}, 
  \label{Eq:HeisenEOM}
\end{eqnarray}
where $\sigma^\pm_\bmk= \frac{1}{2}(\sigma^x_\bmk \pm \ui \sigma^y_\bmk)$, 
$J^L_{\rm sd}(\bmk,\bmq)=\sum_{i \in N/F(L)} J^L_{\rm sd} e^{\ui(\bmk-\bmq)\cdot \bmr_i}$ 
for sublattice $L=A,B$, and we have used the relation
$[\sigma_{\bmk}^{z},\sigma_{\bmk^{'}}^{\pm}] = \pm\frac{2}{\sqrt{N_{N}}}\sigma_{\bmk+\bmk'}^{\pm}$. 
Taking the statistical average of the above quantity, the spin current is calculated to be 
\begin{eqnarray} 
  I_s(t) &=& -\frac{4\sqrt{2}}{\sqrt{N_NN_F} \hbar}
  {\rm Re} \sum_{\bmk, \bmq} 
  \Big[ 
    {\cal J}^+_{\rm sd}(\bmk,\bmq) {\cal A}^<_{\bmk,\bmq}(t,t) \nonumber \\
    && \qquad + 
      {\cal J}^-_{\rm sd}(\bmk,\bmq) {\cal B}^<_{\bmk,\bmq}(t,t)
  \Big], 
\label{Eq:two-Is01}
\end{eqnarray}
where 
${\cal J}^\pm_{\rm sd}(\bmk,\bmq)= 
J^A_{\rm sd}(\bmk,\bmq)\sqrt{S_A}u^{\pm}_{\bmq} + J^B_{\rm sd}(\bmk,\bmq)\sqrt{S_B}u^{\pm}_{\bmq}$. 
Here, ${\cal A}^{<}_{\bmk,\bmq}(t,t') = -\ui \langle \alpha_{\bmq}(t')\sigma_{\bmk}^{-}(t) \rangle$ and 
${\cal B}^{<}_{\bmk,\bmq}(t,t') = - \ui \langle \beta^\dag_{\bmq}(t')\sigma_{\bmk}^{-}(t) \rangle$ 
measure the interface correlation functions between the magnon operators ($\alpha_\bmq$ and $\beta_\bmq$) and spin-density operator $\sigma^-_\bmk$. 
In the steady state the interface correlations ${\cal A}^{<}_{\bmk,\bmq}(t,t')$ and ${\cal B}^{<}_{\bmk,\bmq}(t,t')$ depends only on the time difference $t-t'$. 
Introducing the frequency representation 
${\cal A}^{<}_{\bmk,\bmq}(t,t')= 
\int_{-\infty}^{\infty} \frac{d\omega}{2\pi} {\cal A}^{<}_{\bmk,\bmq}(\omega)e^{-\ui \omega (t-t')}$ as well as using the relationship ${\cal A}^<=\frac{1}{2}[{\cal A}^K-{\cal A}^{R}+ {\cal A}^{A}]$, we finally obtain Eq.~(\ref{Eq:Is-total02}) in the main text. 

\section{Magnon damping near compensation points \label{Sec:Append02}} 

\begin{figure}[t] 
 \begin{center}
  \scalebox{0.3}[0.3]{\includegraphics{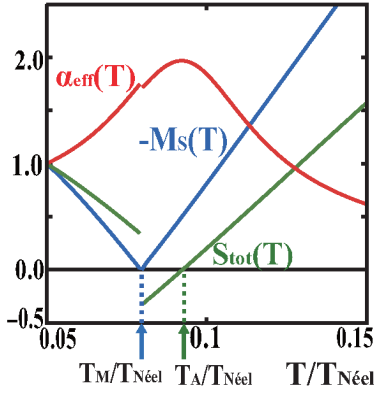}}
  \end{center}
\caption{ (Color online) 
Temperature dependence of the effective magnon damping parameter $\alpha_{\rm eff}$ [red, Eqs.~(\ref{Eq:LL-ferri05}) and~(\ref{Eq:omega01})], saturation magnetization $M_s$ [blue, Eq.~(\ref{Eq:Ms01})], and total angular momentum $S_{\rm tot}$ [green, Eq.~(\ref{Eq:Stot01})], calculated for a compensated ferrimagnet 
Gd$_{23}$Fe$_{74.6}$Co$_{3.4}$. 
The data is normalized by its value at $T/T_{\rm N \text{\'{e}} el }=0.1$. 
}
\label{fig5_Ohnuma}
\end{figure}

In this Appendix, we calculate temperature dependence of magnon damping parameter close to the compensation points and show that the magnon excitation is well defined even at compensation points without any divergences. We begin with two Landau Lifshitz Gilbert equations for sublattice $L=A, B$:~\cite{Stanciu06}  
\begin{eqnarray}
  \frac{d \bmM_L}{dt} &=& -\gamma_L \bmM_L \times \bmH_L + \frac{\alpha_L}{M_{s,L}} \bmM_L \times \frac{d\bmM_L}{dt}, 
\label{Eq:LL-ferri01}
\end{eqnarray}
where 
$ \bmM_L $ is the sublattice magnetization with its magnitude given by $M_L$, 
$\bmH_{L} $ is the effective magnetic field, 
$\gamma_L =g_L \mu_0/\hbar$ is the gyromagnetic ratio, and $\alpha_L $ is the Gilbert damping parameter. 
The effective fields are given by  
$\bmH_A = \bmH_0 + \bmH^{\rm an}_A - \lambda \bmM_B $ 
and $\bmH_B=\bmH_0+\bmH^{an}_B-\lambda \bmM_A$,  
where $\bmH_0=H_0 \widehat{\bmz}$ is external magnetic field, $\bmH^{\rm an}_A= H^{\rm an}_A \widehat{\bmz}$ and $\bmH^{\rm an}_B=- H^{\rm an}_B \widehat{\bmz}$ are the anisotropy fields, 
and $\lambda \bmM_A$ and $\lambda \bmM_B$ are the inter-sublattice exchange fields 
with $\lambda=z_0 J_{AB}/(g_A g_B \mu_0^2)$.  
Because we here focus on the uniform mode, the intra-sublattice exchange couplings 
$\lambda_A=z_0 J_{A}/(g_A \mu_0)^2$ and $\lambda_B=z_0 J_{B}/(g_B \mu_0)^2$ are discarded in Eq.~(\ref{Eq:LL-ferri01}).

Below the magnetization compensation point we set $\bmM_A = M_A \widehat{\bmz} + \bmm_A$ and $\bmM_B = -M_B \widehat{\bmz} + \bmm_B$, such that the effective fields can be written as 
\begin{eqnarray}
  \bmH_A &=& ( H_0 + H^{\rm an}_A + \lambda M_{B} ) \widehat{\bmz} - \lambda \bmm_B, \\
  \bmH_B &=& (H_0 - H^{\rm an}_B - \lambda M_{A} ) \widehat{\bmz} - \lambda \bmm_A. 
\label{Eq:Heff01}
\end{eqnarray}
Introducing the representation $E_A^{\rm eff}= - (H_0 + H_A^{\rm an} + \lambda M_B)$ and 
$E_B^{\rm eff}= -(H_0 - H_B^{\rm an} - \lambda M_A)$, and linearizing with respect to $\bmm_A$ and $\bmm_B$, the Landau-Lifshitz-Gilbert equations are transformed to be 
\begin{eqnarray}
  \frac{d \bmm_A}{dt} &=& \widehat{\bmz} \times \Big[ 
    \gamma_A (\lambda M_A \bmm_B -E_A^{\rm eff} \bmm_A) 
    + \alpha_A \frac{d \bmm_A}{dt} \Big], \nonumber \\
  \label{Eq:LL-ferri03a} \\
  \frac{d \bmm_B}{dt} &=& - \widehat{\bmz} \times \Big[
  \gamma_B (\lambda M_B \bmm_A +E_B^{\rm eff} \bmm_B) 
  + \alpha_B \frac{d \bmm_B}{dt} \Big]. \nonumber \\
\label{Eq:LL-ferri03b}
\end{eqnarray}
We introduce $m^\pm= m_x \pm \ui m_y$ and substitute $m^+_{L}(t)= m^+_{L}e^{-\ui \omega t}$ into Eqs.~(\ref{Eq:LL-ferri03a}) and~(\ref{Eq:LL-ferri03a}). Then we obtain 
\begin{eqnarray}
  (-\ui \omega-\alpha_A \omega + \ui \gamma_A E_A^{\rm eff}) 
  m^+_A - \ui \lambda \gamma_A M_A m^+_B &=& 0, \nonumber \\
  \\
  (- \ui \omega+\alpha_B\omega +i\gamma_B E_B^{\rm eff}) 
  m^+_B + \ui \lambda \gamma_B M_B m^+_A &=& 0. \nonumber \\
\label{Eq:LL-ferri04}
\end{eqnarray}
The eigenfrequency $\omega$ is determined by the equation: 
\begin{eqnarray}
  &&(\omega - \ui \alpha_A \omega - \gamma_{A} E_{A}^{\rm eff})
  (\omega+ \ui \alpha_B \omega - \gamma_{B} E_{B}^{\rm eff}) \nonumber \\
  &&+ \lambda^{2} \gamma_{A}\gamma_{B} M_{A}M_{B} = 0. 
\label{Eq:LL-ferri05}
\end{eqnarray}

Above the magnetization compensation temperature, we set $\bmM_A=-M_A \widehat{\bmz}+\bmm_A$ and $\bmM_B=M_B \widehat{\bmz}+\bmm_B$ because we consider a situation in which the saturation magnetization is pinned by an external magnetic field. 
This situation can be analyzed by rewriting $E_A^{\rm eff}=-(H_0 - H_A^{\rm an} - \lambda M_B)$ 
and $E_B^{\rm eff}=-(H_0 + H_A^{\rm an} + \lambda M_A)$ as well as reversing the signs of $\alpha_{A}$ and $\alpha_{B}$. 
We numerically solve Eq.~(\ref{Eq:LL-ferri05}) by setting 
\begin{equation}
  \omega= \omega_{0}+\ui \alpha_{\rm eff} \omega_{0}. 
  \label{Eq:omega01}
\end{equation}

Figure~\ref{fig5_Ohnuma} shows the temperature dependence of the effective Gilbert damping parameter $\alpha_{\rm eff}$ the lower frequency mode, calculated for a compensated ferrimagnet Gd$_{22}$Fe$_{70}$Co$_8$.~\cite{Stanciu06,Taylor77,Hansen89} We assign $A$ sublattice for Gd ions and $B$ sublattice for Fe ions, and neglect Co ions for simplicity. We set $S_{A}=3.85$, $S_{B}=3.5$, $g_{A}=2.0$, $g_{B}=2.05$, $H_{0}=0.04$~T, $H_{A}^{\rm an}= 0.0$~T, $H_{B}^{\rm an} =0.02$~T, $\alpha_A=0.004$, $\alpha_B=0.0039$. 
The saturation magnetization and the total spin are calculated by using the mean field approximation by using $J_{AB}=0.28$~meV, $J_A=0$~meV, and $J_B=0.34$~meV to reproduece $T_{\rm N \text{\'{e}} el }=500$~K. These parameters give $T_{\rm M} \approx 0.079 T_{\rm N \text{\'{e}} el }$ and $T_{A} \approx 0.091 T_{\rm N \text{\'{e}} el }$. From the figure, we see that Gilbert damping constant is enhanced around the angular momentum compensation point $T_{\rm A}$, but does not show any divergences. 
A small discontinuity at $T_{\rm M}$ stems from the fact that the spin quantization axis is reversed at $T_{\rm M}$ because of the pinning by the external magnetic field. 
The result obtained here justifies the statement in Sec.~\ref{Sec:Conclusion} that the presumed divergence of the magnon damping parameter at $T_{\rm A}$~\cite{Stanciu06} does not exist when we recall the condition justifying the expansion used in Ref.~\onlinecite{Wangsness53}, where such an effect manifests itself as an enhancement of the damping parameter without any divergence.

\end{document}